\documentstyle[aps,12pt,float,epsfig,amssymb]{revtex}

\begin{document}
\baselineskip=0.7cm
\newcommand{\ini}{\begin{equation}}
\newcommand{\fin}{\end{equation}}
\newcommand{\inir}{\begin{eqnarray}}
\newcommand{\finr}{\end{eqnarray}}
\newcommand{\inif}{\begin{figure}}
\newcommand{\finf}{\end{figure}}
\newcommand{\bc}{\begin{center}}
\newcommand{\ec}{\end{center}}
\def\ol{\overline}
\def\pa{\partial}
\def\ra{\rightarrow}
\def\ts{\times}
\def\df{\dotfill}
\def\bs{\backslash}
\def\dg{\dagger}

$~$

\hfill DSF-07/2001

\vspace{1 cm}

\centerline{{\bf NEUTRINO OSCILLATIONS AND}} 
\centerline{{\bf NEUTRINOLESS DOUBLE BETA DECAY}}

\vspace{1 cm}

\centerline{\large{D. Falcone and F. Tramontano}}

\vspace{1 cm}

\centerline{Dipartimento di Scienze Fisiche, Universit\`a di Napoli,}
\centerline{Complesso di Monte Sant'Angelo, Via Cintia, Napoli, Italy}
\centerline{e-mail: falcone@na.infn.it}

\vspace{1 cm}

\begin{abstract}

\noindent
The relation between neutrino oscillation parameters and
neutrinoless double beta decay
is studied, assuming normal and inverse hierarchies for Majorana neutrino
masses. For normal hierarchy the crucial dependence on $U_{e3}$ is explored.
The link with tritium beta decay is also briefly discussed.
 
\end{abstract}

\newpage

\noindent
There is now convincing evidence for neutrino masses and lepton mixings from
oscillation experiments. Neutrino masses can be either of the Dirac type or
of the Majorana type. In the case of Majorana masses the neutrinoless double
beta decay ($0 \nu \beta \beta$ decay) is allowed \cite{furry}.
Such a decay has not yet been observed and only an upper 
limit on the related mass parameter $M_{ee}$ (to be defined below) is available,
$M_{ee} < 0.2$ eV \cite{mee}
(in this paper $M_{ee}$ is always expressed in eV). 
Several experiments have been proposed to lower this limit by one or even two
orders and eventually discover the $0 \nu \beta \beta$ decay, thus revealing
the Majorana nature of neutrinos. Therefore, the subject of the relation
between oscillation parameters and $M_{ee}$ has been studied by many authors
\cite{kps,gbgkm,many}. Here we turn to the question in order to clarify the
link between the lepton mixings and the predictions for $M_{ee}$.

In fact, the random extraction of the relevant neutrino parameters is
very useful in this case. In particular, we will see that for
$U_{e3} \lesssim 0.05$ the four solutions to the solar neutrino problem
may give quite different predictions for the mass parameter $M_{ee}$.
Since also the
bound $U_{e3} < 0.2$ \cite{chooz} is expected to be lowered in the future,
phenomenological relations between $M_{ee}$ and $U_{e3}$ are welcome.
We consider normal and inverse hierarchies for the Majorana masses of three
active neutrinos. Recent evidence for cosmological dark energy eliminates
most of the motivations for considering the degenerate spectrum, which was
before relevant for hot dark matter (see for example \cite{lan}).

Let us now define the mass parameter $M_{ee}$ as
\ini
M_{ee}= |U_{e1}^2 e^{2 \text{i} \alpha} m_1+  
U_{e2}^2 e^{2 \text{i} \beta} m_2+ U_{e3}^2 m_3|,
\fin
where $U_{ei}$ ($i=1,2,3$) are the moduli of the elements in the first row
of the lepton mixing matrix $U$.
This matrix can be parametrized as the standard form of the CKM matrix
(with one phase $-\delta$ in entry 1-3) times
$\text{diag} (\text{e}^{\text{i} \varphi_1},\text{e}^{\text{i} \varphi_2},1)$.
Two relative phases $\alpha=\varphi_1+\delta$ and $\beta=\varphi_2+\delta$
appear in $M_{ee}$. Moreover, $m_i$ are positive
Majorana masses. From ref.\cite{bg} we get
\ini
U_{e2}^2=(\sin^2 \theta_s)(1-U_{e3}^2),
\fin
where $\theta_s$ is the solar neutrino mixing angle, and due to the
unitarity of $U$ we have 
\ini
U_{e1}^2=1-U_{e2}^2-U_{e3}^2=(\cos^2 \theta_s)(1-U_{e3}^2).
\fin
Therefore, neglecting $U_{e3}^2$ with respect to 1, eqn.(1) can be written as
\ini
M_{ee}=|(\cos^2 \theta_s)e^{2 \text{i} \alpha} m_1+
(\sin^2 \theta_s)e^{2 \text{i} \beta} m_2+U_{e3}^2 m_3|.
\fin
In this way $M_{ee}$ depends on seven neutrino parameters. As said above the
mixing $U_{e3}$ is bounded (by the CHOOZ experiment),
\ini
U_{e3}<0.2.
\fin
In order to determine the masses $m_i$ we have to distinguish between the
normal mass hierarchy, $m_1 \ll m_2 \ll m_3$, and the inverse mass hierarchy,
$m_1 \simeq m_2 \gg m_3$. In the normal hierarchy case
\ini
m_3=\sqrt{\Delta m^2_a+m_1^2},
\fin
\ini
m_2=\sqrt{\Delta m^2_s+m_1^2},
\fin 
and for $m_1$ we take
$10^{-5} \sqrt{\Delta m^2_s} <m_1< 10^{-1} \sqrt{\Delta m^2_s}$, with
$\Delta m^2_a =(1-6) \ts 10^{-3} \text{eV}^2$ for atmospheric neutrinos, and
$\Delta m^2_s$ reported in Table I (in eV$^2$) together with $\sin \theta_s$
for solar neutrinos. These values come from ref.\cite{bks}.
LMA, SMA, LOW are the large mixing angle, small mixing angle, low mass
matter (MSW) solutions, and VO is the vacuum solution. The best global fit
of solar neutrino data is given by the LMA solution, although the other
solutions are not ruled out \cite{lan}.
The value $\sin \theta_s=0.71$ ($\theta_s =\pi/4$) means maximal mixing.
For inverse hierarchy one has
$m_1 \simeq m_2 \simeq \sqrt{\Delta m^2_a}$. 

Let us consider first the normal hierarchy. The results of the calculation
(2500 points extracted) are in Figs. 1-2, where we plot
$\text{Log}_{10} M_{ee}$ versus $U_{e3}$. For $U_{e3}>0.1$ the SMA, LOW and VO
solutions give similar values for $M_{ee}$, while the LMA solution provides
also higher values.
However, the LMA solution gives $M_{ee}$ almost constant,
because the $m_3$-term is negligible even for $U_{e3} \simeq 0.2$,
while for the other solutions $M_{ee}$ decreases for smaller $U_{e3}$, till
the $m_3$-term becomes negligible. In particular,
for $U_{e3} \lesssim 0.05$ the LMA solution is clearly distinguished from the
SMA solution (and also from the others). A similar behaviour happens for the LOW
solution with respect to the VO solution, for $U_{e3} \lesssim 0.02$.
In order to clarify this aspect we have checked the results on a linear plot.
In Fig. 3 we report the lower LMA bound and the upper SMA bound as well as
the lower LOW bound and the upper VO bound for $M_{ee}$.
The lower LMA bound can be obtained from the expression
$M_{ee} \simeq m_2 \sin^2 \theta_s-m_3 U_{e3}^2$ and the upper SMA bound
from $M_{ee} \simeq m_1+m_3 U_{e3}^2$.
We can thus predict, for $U_{e3} \lesssim 0.05$, that
$4 \ts 10^{-4} < M_{ee} < 8 \ts 10^{-3}$ for the LMA solution, while 
$M_{ee} < 4 \ts 10^{-4}$ for the SMA solution (and also the LOW solution).
For $U_{e3} \lesssim 0.01$ we get $3 \ts 10^{-5} < M_{ee} < 4 \ts 10^{-4}$
for the LOW solution, while $M_{ee} < 3 \ts 10^{-5}$ for the VO solution.
We now comment about the cancellations appearing in Figs. 1-2. They are
obtained  when the $m_2$-term and/or the $m_1$-term are comparable with the
$m_3$-term. Of course, this happens for different $U_{e3}$ values
(and the relevant phase tuned around $\pi/2$),
according to the different solar solutions. Note
also that in the SMA case the $m_1$-term can easily exceed the $m_2$-term.
In the other cases, for $U_{e3} \simeq 0$, we have
$M_{ee} \simeq (\sin^2 \theta_s)m_2 \simeq (\sin^2 \theta_s)
\sqrt{\Delta m^2_s}$.

For the inverse hierarchy two main results can be drawn out.
One is with respect to the normal hierarchy, namely in the region
$10^{-2} < M_{ee} < 10^{-1}$ only the inverse hierarchy is possible, while
$M_{ee} < 10^{-2}$ for the normal hierarchy. The other result is that the
SMA solution gives clean bounds, $3 \ts 10^{-2} < M_{ee} < 8 \ts 10^{-2}$.
All solutions have the same upper bound $M_{ee} < 8 \ts 10^{-2}$, not depending
on $U_{e3}$, which is easily understood since the $m_3$-term is negligible for
inverse hierarchy. The basic features of
the inverse hierarchy case can be obtained by using the approximation
\ini
M_{ee} \simeq
(\cos^2 \theta_s \pm \sin^2 \theta_s) m_{1,2}
\fin
in the CP-conserving case ($\alpha=0$, $\beta=0,\pi/2$). In fact, the plus sign
($\beta=0$) gives $M_{ee} \simeq m_{1,2} \simeq \sqrt{\Delta m^2_a}$,
while the minus sign ($\beta=\pi/2$)
gives $M_{ee} \simeq \sqrt{\Delta m^2_a} \cos 2 \theta_s$. For small mixing
$\theta_s \simeq 0$ one has $M_{ee} \simeq \sqrt{\Delta m^2_a}$, while for
large mixing $\theta_s \simeq \pi/4$ the value $M_{ee} \simeq 0$
is allowed by cancellations,
so that the full range $0 \le M_{ee} \le \sqrt{\Delta m^2_a}$ is covered.
Of course, if maximal mixing is excluded, a lower bound appears also for the
LMA, LOW and VO solutions.

Now we discuss the mass parameter $m_{\beta}$, related to tritium beta decay,
which is defined as
\ini
m_{\beta}^2= U_{e1}^2 m_1^2+U_{e2}^2 m_2^2+U_{e3}^2 m_3^2.
\fin
Using eqns.(2),(3) and neglecting again $U_{e3}^2$ with respect to 1, we obtain
\ini
m_{\beta}^2= (\cos^2 \theta_s) m_1^2+(\sin^2 \theta_s) m_2^2+U_{e3}^2 m_3^2.
\fin
There are no cancellations for $m_{\beta}$, so that it is sufficient
to evaluate $M_{ee}$ in $U_{e3} \simeq 0$ and $U_{e3} \simeq 0.2$.
In the normal hierarchy case, for $U_{e3} \simeq 0.2$ we get
$m_{\beta} \simeq U_{e3} m_3 \simeq U_{e3} \sqrt{\Delta m^2_a}$.
For $U_{e3} \simeq 0$ the SMA solution gives
$m_{\beta} \simeq m_1 \ll \sqrt{\Delta m^2_s}$, while the other solutions give 
$m_{\beta} \simeq (\sin \theta_s)m_2 \simeq
(\sin \theta_s) \sqrt{\Delta m^2_s}$.
For inverse hierarchy all solutions give
$m_{\beta} \simeq m_{1,2} \simeq \sqrt{\Delta m^2_a}$,
not depending on $U_{e3}$.
The experimental limit on $m_{\beta}$ is now $m_{\beta} < 2.2$ eV
(see \cite{akh}) and it is hard to lower this limit by one order.
However, the maximum value allowed by the previous discussion is
$m_{\beta} \simeq 8 \ts 10^{-2}$ eV, so that the impact of neutrino oscillations
on the prediction for $m_{\beta}$ cannot be checked.

In conclusion, we have studied the prediction for $M_{ee}$ obtained by
varying neutrino parameters, within the experimental ranges,
for the normal and inverse mass hierarchy cases.
For normal hierarchy the main result is that for
$U_{e3} \lesssim 0.05$ the LMA solution is clearly distinguished from the
other solutions. Moreover, for $U_{e3} \lesssim 0.01$ the LOW solution is
distinguished from the VO solution. This means that if the LMA solution is
confirmed, and even if $U_{e3}$ is very small (similar to $V_{cb}$ or $V_{ub}$),
the GENIUS II (10 t) experiment should find the $0 \nu \beta \beta$ decay
unless neutrinos are Dirac particles. Instead, if another solution is confirmed
and $U_{e3} \lesssim 0.1$, then the GENIUS project will not be able to decide
about the neutrino nature.
For inverse hierarchy $M_{ee}$ could
be higher by one order, with respect to normal hierarchy,
and the $0 \nu \beta \beta$ decay be possibly
found also by the GENIUS I (1 t) and MOON experiments.

In this paper we have taken $0 < \theta_s \le \pi/4$. However, for LOW and VO
solutions, part of the range $\pi/4 < \theta_s < \pi/2$ (the so-called
dark side of neutrino parameter space \cite{dark}) is allowed
(see for example \cite{mon}), so that for normal hierarchy the
related regions in $M_{ee}$ can overlap also for $U_{e3} \simeq 0$. Of course,
progress in the determination of neutrino oscillation parameters will sharpen
the predictions on $M_{ee}$ for both hierarchies. 

\newpage

\begin{table}
\caption{Neutrino oscillation parameters}
\begin{tabular}{ccccc}
$~$ & LMA & SMA & LOW & VO \\ \hline 
$\Delta m^2_s$ & $(0.15-1.5) \ts 10^{-4}$ & $(0.4-1) \ts 10^{-5}$ &
$(0.3-2.5) \ts 10^{-7}$ & $(0.3-10) \ts 10^{-10}$ \\
$\sin \theta_s$ & $0.40-0.71$ & $0.02-0.05$ & $0.53-0.71$ & $0.43-0.71$
\end{tabular}
\end{table}

\newpage

\begin{figure}[ht]
\begin{center}
\epsfig{file=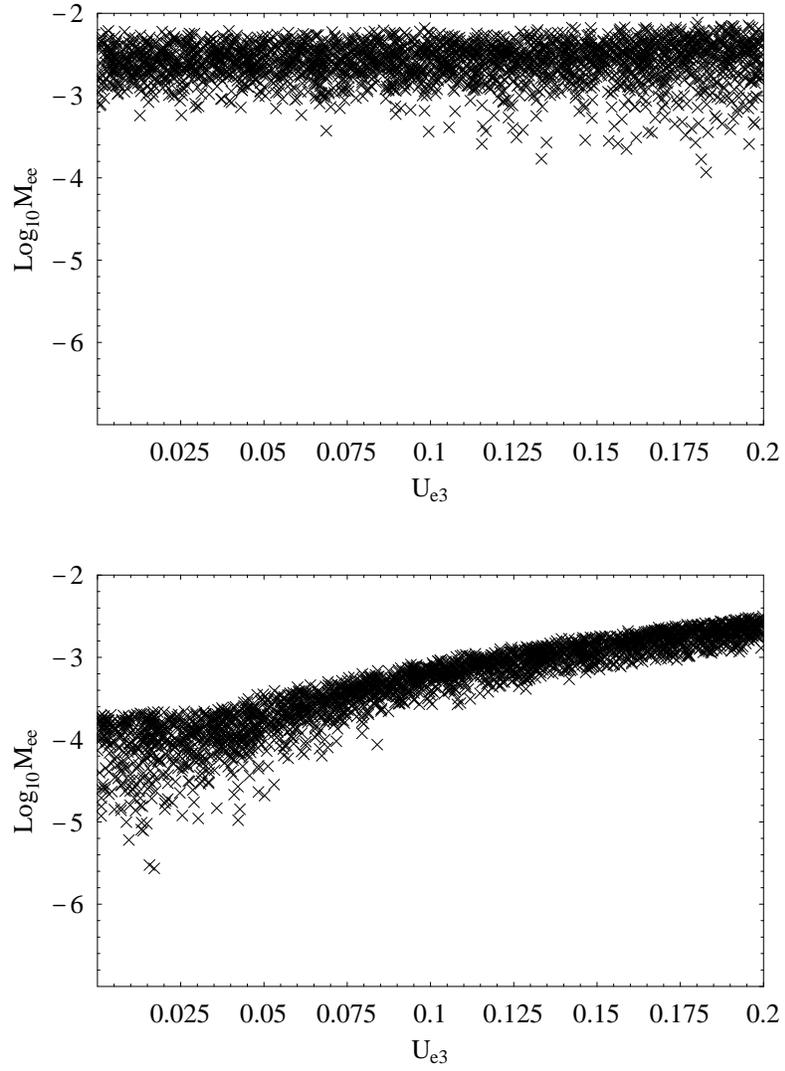,height=15cm}
\caption{$\text{Log}_{10} M_{ee}$ vs $U_{e3}$ for the LMA and SMA solutions
with the normal hierarchy}
\end{center}
\end{figure}

\newpage

\begin{figure}[ht]
\begin{center}
\epsfig{file=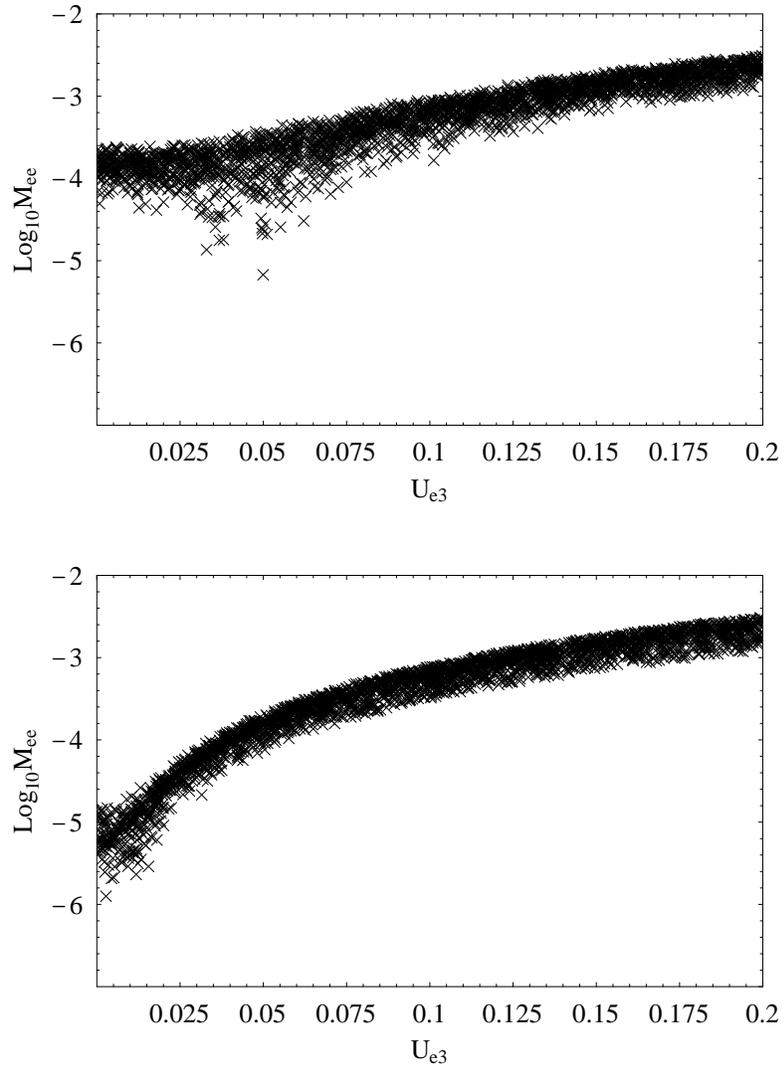,height=15cm}
\caption{$\text{Log}_{10} M_{ee}$ vs $U_{e3}$ for the LOW and VO solutions
with the normal hierarchy}
\end{center}
\end{figure}

\newpage

\begin{figure}[ht]
\begin{center}
\epsfig{file=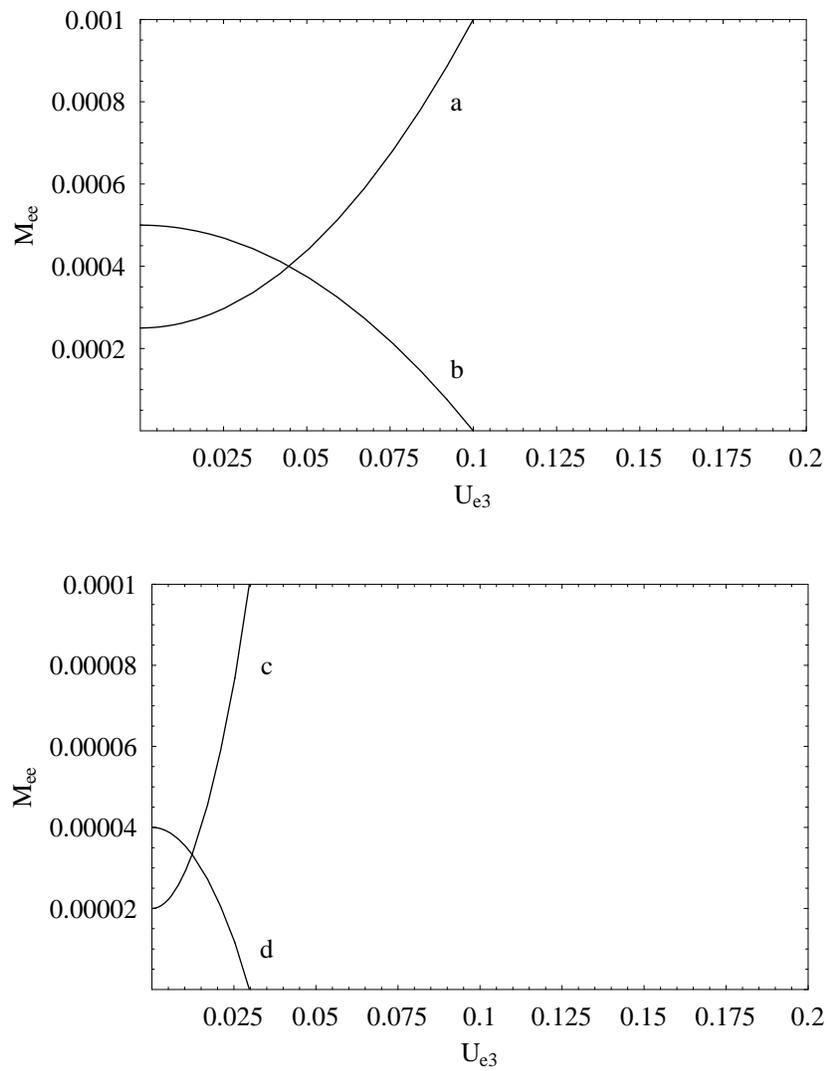,height=15cm}
\caption{Upper SMA bound (a) and lower LMA bound (b) as well as
upper VO bound (c) and lower LOW bound (d) for $M_{ee}$ with the
normal hierarchy}
\end{center}
\end{figure}

\end{document}